\documentclass[amsmath,amssymb,nofootinbib,prd]{revtex4}

\usepackage{setspace}
\usepackage{graphicx}

\usepackage{amsmath,amssymb,amsfonts,amsthm}
\usepackage{enumerate}

\begin{document}

\title{Smooth extensions of black holes  in loop quantum gravity}

\author{Rodolfo Gambini$^1$, Javier Olmedo$^2$,  Jorge Pullin$^3$}
\affiliation{1. Instituto de F\'{\i}sica, Facultad de Ciencias, Igu\'a 4225, esq. Mataojo,
11400 Montevideo, Uruguay. \\
2. Departamento de F\'isica Te\'orica y del Cosmos, Universidad de Granada,  Granada-18071, Spain\\
3. Department of Physics and Astronomy, Louisiana State University,
Baton Rouge, LA 70803-4001, USA.}

\begin{abstract}
Vacuum spherically symmetric loop quantum gravity in the midi-superspace approximation using inhomogeneous horizon-penetrating slices has been studied for a decade, and it has been noted that the singularity is eliminated. It is replaced by a region of high curvature and potentially large quantum fluctuations. It was recently pointed out that the effective semiclassical  metric  implies the existence of a shell of effective matter which violates energy conditions in regions where the curvature is largest. Here we propose an alternative way of treating the problem that is free from the shells. The ambiguity in the treatment is related with the existence of new observables in the quantum theory that characterize the area excitations, and how the counterpart of diffeomorphisms in the discrete quantum theory is mapped to the continuum semi-classical picture.
The resulting space-time in the high curvature region inside the horizon is approximated by a metric of the type of the Simpson--Visser wormhole and it connects the black hole interior to a white hole in a smooth manner.
\end{abstract}
\maketitle

\section{Introduction}

Studies of spherically symmetric space-times have been analyzed in loop quantum gravity using a midi-superspace approximation for about a decade. They have been carried out using both inhomogeneous slices that cover both the interior and exterior \cite{review}, in the case of space-times with horizons, and also using the isometry between the Schwarzschild interior and the Kantowski--Sachs cosmologies \cite{lqc}. In all cases it is observed that the singularity is eliminated. It gets replaced by a region of space-time of high curvature and possibly large quantum fluctuations, that an observer can in principle traverse into a region towards the future, giving rise to a white hole. It has been argued that   the latter might lead to observable consequences \cite{fireworks}. It has recently been noted \cite{shells} that the inhomogeneous treatments involve an extension of the effective metric to negative values of the $r$ coordinate that has discontinuous derivatives. As a consequence, a shell of matter arises at the most quantum region, and it violates energy conditions. The latter is not a surprise: the effective geometry is in general non-vacuum and associated with effective matter violating energy conditions. That is how the elimination of singularities can be understood from the point of view of the semi-classical theory. It should be noted that the solutions are vacuum solutions, the matter that arises is effective and can be viewed as emerging due to quantum effects.

It turns out that the way the solutions are constructed at the most quantum region is not unique. This is tantamount to the choice of semiclassical states describing the system. In this paper we identify suitable semiclassical states that yield smooth extensions of the metric through the most quantum region that do not give rise to shells of matter there.
The resulting space-time at the throat is isometric to the Simpson--Visser wormhole with the throat inside the horizon, and it connects smoothly the interior of a Schwarzschild black hole to a white hole in the future.

The organization of this paper is as follows: in section 2 we review the traditional treatment, in section 3 we present the new approach and in section 4 we discuss the metric and the effective stress tensor. We end with conclusions.

\section{Quantum theory: traditional treatment}

Spherically symmetric space-times are described in terms of Ashtekar variables by a pair of triads, one in the radial direction $E^x$ and one in the tangential 
one $E^\varphi$ and their canonically conjugate momenta $K_x$ and $K_\varphi$. These variables bear a simple relation with the traditional metric variables, we refer the reader to our review \cite{review} for this and other details. The quantum states are constructed using one dimensional holonomies along the radial direction of the variable $K_x$ between the nodes of a graph and point holonomies for the variable $K_\varphi$ at the nodes.

The Hamiltonian  constraint can be abelianized by redefining the lapse and shift, and the physical space of states can be explicitly determined. The solutions  are labeled by a vector of positive integers ${\vec k}=(k_1,k_2,\ldots k_V)$ where each $k_i$ represents the area of a sphere of symmetry with radial coordinate $x_i$ such that $x_i=\sqrt{k_i}\ell_{\rm Planck}$ and a real  variable $M$ representing the ADM mass. The physical states are given by one-dimensional ``spin networks" $\vert {\vec k},M\rangle$.

Let us recall how the usual construction (the one that gives rise to shells) works. For each state in the physical space one can define the action of the radial triad and its derivative as,
\begin{align}\label{eq:hex}
  &|\hat E^x(x_j)||\vec k,M\rangle=\hat O(z(x_j))|\vec k,M\rangle=\ell_{\rm Pl}^2 k_{j}|\vec k,M\rangle=x^2_{j}|\vec k,M\rangle.\\\label{eq:hdex}
  & [\hat E^x(x_j)]'|\vec k,M\rangle=\ell_{\rm Pl}^2\frac{k_{j+1}-k_j}{\delta x_j}|\vec k,M\rangle=\frac{(x_j+{\delta x}_j)^2-x_j^2}{\delta x_j}|\vec k,M\rangle={\rm sign}(j)(2x_j+{\delta x_j})|\vec k,M\rangle.
\end{align} 
In this construction $z(x_j)$ is a function that corresponds to a choice of the radial coordinate. That makes the operators defined in the above equations (parameterized) Dirac observables. They are operators that have vanishing commutators with the constraints but depend on a (functional) parameter. Until one chooses it, they do not take a definite value. They can be viewed as gauge dependent quantities evaluated in a gauge; that is how kinematical quantities like $E^x$ can be thought of as Dirac observables, and so is the metric of space-time. For a review of this construction we refer the reader to \cite{review}. In terms of them, the metric reads,
\begin{eqnarray}
\hat g_{tt}(x_j) &=& -\left(1-\frac{\hat r_S}{\sqrt{\hat |E^x(x_j)|}}\right),\\\label{3}
\hat g_{tx}(x_j) &=& -\sqrt{\frac{\pi}{\Delta}}\frac{\left\{\widehat{\left[E^x(x_j)\right]'}\right\}{\sqrt{[\hat F(x_j)]^2}}}{\sqrt{1-\frac{\hat r_S}{\sqrt{|\hat E^x(x_j)|}}+\frac{4\pi |\hat E^x(x_j)| [\hat F(x_j)]^2}{\Delta}}},\\
\hat g_{xx}(x_j) &=& \frac{\left\{\widehat{\left[E^x(x_j)\right]'}\right\}^2}{4 |\hat E^x|\textbf{}\left(1-\frac{\hat r_S}{\sqrt{|\hat E^x(x_j)|}}+\frac{4\pi |\hat E^x(x_j)| [\hat F(x_j)]^2}{\Delta}\right)},\\
\quad\hat g_{\theta\theta}(x_j)&=&|\hat E^x(x_j)|,\quad \hat g_{\phi\phi}(x_j)=|\hat E^x(x_j)|\sin^2\theta,\label{6}
%\label{eq:hatgmunu2}
\end{eqnarray}
where $\Delta$ is the area gap in loop quantum gravity, usually taken as proportional to $\ell_{\rm Planck}^2$.

To compare with the usual classical general relativity results in terms of a metric geometry in previous works  we made some additional assumptions.
We  considered the leading quantum corrections when the dispersion of the mass can be neglected compared to the spacing of the discretization implied by the spin network. This allows us to proceed to drop all hats in the above expression and call the result ${}^{(0)}g_{\mu\nu}(x_j)$. We also took, for convenience, states such that 
\begin{equation}
x_j=|j| \delta x +x_0    
\end{equation}
 where we have taken $\delta x$ to be an integer times  $\ell_{\rm Planck}$. This ensures that $k_j$ takes integer values. So we chose an equally spaced spin network.

The improved quantization conditions discussed in \cite{improved} imply that there is a minimum value of $k_j$, that we call $k_0$, implied by the quantization of area. This in turn leads to a minimum value of $x_j$ given by,
 $x_0=\ell_{\rm Pl}\,{\rm Int}\left[\sqrt{k_0}\right]$ with $k_0$ 
\begin{equation}
  k_0 = {\rm Int}\left[\left(\frac{2 G M \Delta}{4\pi \ell_{\rm Pl}^3}\right)^{2/3}\right].\label{7}
\end{equation}

We approximate the expectation value of the discrete  metric operator defined in (\ref{3}-\ref{6}) by
keeping terms $\delta x/x_j$ and  writing them as $\delta x/(|x|+x_0)$ at first order. This makes explicit that the effective geometries ``bounce'' when they reach $x=0$. 

The functions $\hat{F}(x_j)$ determine the slicing condition. For instance, for Painlev\'e--Gullstrand \cite{review} we have that,
\begin{align}\label{eq:gf-f1}
\hat F(x_j)=\bar\rho_j\sqrt{\frac{\hat r_S}{\sqrt{\hat E^x(x_j)}}}.
\end{align}
and the effective semiclassical metric takes the form,
\begin{eqnarray}
{}^{(0)}  g_{tt}(x) &=&
\label{eq:hatgmunu31}-\left(1-\frac{r_S}{|x|+x_0}\right),\\
{}^{(0)}  g_{tx}(x) &=& -{\rm sign}(x)\sqrt{\frac{r_S}{|x|+x_0}}\left(1+\frac{\delta x}{2(|x|+x_0)}\right)\,,\\
{}^{(0)}  g_{xx}(x) &=& \left(1+\frac{\delta x}{2(|x|+x_0)}\right)^2, \quad {}^{(0)} g_{\theta\theta}(x)=(|x|+x_0)^2,\\ {}^{(0)}g_{\phi\phi}(x)&=&(|x|+x_0)^2\sin^2\theta.\label{eq:hatgmunu3}
\end{eqnarray}

We see the emergence of the absolute values that allow to extend the metric to negative values of the radial coordinate. However, this means that the metric  in Painlev\'e--Gullstrand coordinates has a discontinuous component ${}^{(0)}g_{tx}$ (that issue may disappear in other coordinates) and discontinuous derivatives. The discontinuity in the derivatives is always present, independently of the coordinates chosen. It suffices to choose $x_j=\Delta x_j+x_0$ with $\Delta x_j=\ell_{\rm Planck}\sqrt{k_j}-x_0$. It is the parameterization at the neighborhood of the bounce  what gives rise to the shell of matter at $x=0$. It should be noted that in all cases in the limit $\hbar\to 0$ one recovers the Schwarzschild solution everywhere with the usual singularity at $x=0$.

The reader may question the meaning of discontinuities in a theory that is discrete to begin with. However, given that one can choose quantum states where the separation of the spin network nodes is only constrained by the quantization of areas of symmetry ---which yields a bound considerably smaller than Planck's length---, the solutions can approximate the continuum very well. Yet, the discontinuities in question persist in that limit and are large.

\section{Quantum theory: new description of the bounce}

We now introduce a new description of the bounce, free of discontinuities.
For this purpose, we will consider a new family of quantum states that also takes values for negative $k$'s.  For simplicity, we will choose a particular quantum state to present the idea; it can be extended to any physical state as we shall see later on. Our choice is,
\begin{align}
%\label{eq:hex2}
  &\hat E^x(x_j)|\vec k,M\rangle=\hat O(z(x_j))|\vec k,M\rangle=\ell_{\rm Pl}^2 k_{j}|\vec k,M\rangle={\rm sign}(k_j)(x^2_{j}+x_0^2)|\vec k,M\rangle,
  \label{eq:hex2}
\end{align} 
where $k_j$ is an arbitrary integer, and we have chosen a spacing of the nodes given by
$\delta x=\ell_{\rm Pl}\, {\rm Int}[x_0/\ell_{\rm Pl}]$  for convenience (though there is some freedom in this choice) and define the new radial coordinate as 
\begin{equation}
x_j=j\delta x,
\end{equation} 
with $j>0$. We will adopt the same symbol $x_j$ in the following.  On the other hand, the operator $\hat [E^x(x_j)]'$ has eigenvalues given by the first equality in equation (\ref{eq:hdex}). But, on the states given by Eq. \eqref{eq:hex2}, its spectrum is positive definite (its minimum eigenvalue equals $\delta x$). As we discussed in the previous section, in past papers we choose a different state 
at the region where the singularity appears in the classical theory.  Just like in the previous section, the (smeared) triads along the radial direction are multiplicative operators, and so are their spatial derivatives and
in terms of them, one can write the components of the metric as a Dirac observable dependent on (functional) parameters,
\begin{eqnarray}
\hat g_{tt}(x_j) &=& -\left(1-\frac{\hat r_S}{\sqrt{|\hat E^x(x_j)|}}\right),\label{15}\\
\hat g_{xx}(x_j) &=& \frac{\left\{\widehat{E^x(x_j)'}\right\}^2}{4 |\hat E^x(x_j)|\textbf{}\left(1-\frac{\hat r_S}{\sqrt{|\hat E^x(x_j)|}}+\frac{4\pi |\hat E^x(x_j)| [\hat F(x_j)]^2}{\Delta}\right)},\\
\quad\hat g_{\theta\theta}(x_j)&=&|\hat E^x(x_j)|,\quad \hat g_{\phi\phi}(x_j)=|\hat E^x(x_j)|\sin^2\theta,\label{17}
%\label{eq:hatgmunu2}.
\end{eqnarray}

As before, since the spin networks are eigenstates of the triads, it is straightforward to evaluate the action of the metric parameterized Dirac observable on them once a foliation given by $F(x_j)$ is chosen and the corresponding lapse and shift determined. In the Schwarzschild gauge $F=0$ and neglecting the corrections due to $\delta x$, the approximate semi-classical metric takes the form,
\begin{equation}
ds^{2}=-\left(1-\frac{2GM}{\sqrt{x^{2}+x_0^{2}}}\right)dt^{2}+\frac{dx^{2}}{1-\frac{2 G M}{\sqrt{x^{2}+x_0^{2}}}}
+\left(x^{2}+x_0^{2}\right)\left(d\theta^{2}+\sin^{2}\theta \;d\phi^{2}\right),\label{17b}
\end{equation}
where $x_0^2=\ell_{\rm Pl}^2\,k_0$, with $k_0$ given in Eq. \eqref{7}. 

For $x_0>2GM$ this metric represents the Simpson--Visser \cite{SimpsonVisser} wormhole, a particular case of those studied by Morris and Thorne \cite{MorrisThorne}. The Simpson--Visser geometry is Ricci flat, and regular at the origin, unless $x_0\to 0$. Ours is a case with $x_0\ll 2GM$, but with $x_0> 0$. The lower and positive bound of the parameter $x_0$ is determined by the theory. Even though it is small, it is a positive quantity that characterizes the size of the throat at the bounce of the black hole into a white hole.

\section{The metric and effective stress tensor}

To compute the effective stress-energy tensor of the metric we need to include the terms in the expansion in terms of the spacing $\delta x$ that were neglected in (\ref{17b}). 
Since the spin network states are eigenstates of the triads, it is straightforward to evaluate the action of the parameterized Dirac observable corresponding to the metric once a foliation $F(x_j)$ is chosen and the corresponding lapse and shift determined. As we mentioned before, we consider a sufficiently peaked superposition of states of the physical Hilbert space with $k_{0}<\tilde{k}_0(M)<k_0+1$, with $\tilde{k}_0(M)$  given by (\ref{7})
and $x_0=\ell_{\rm Pl}\,\sqrt{k_0}$. Elements of the physical space of states $\vert \vec{k},M\rangle$ satisfy (\ref{eq:hex2}) and the normalization condition $\langle \vec{k},M,\vert \vec{k}',M'\rangle=\delta(M-M')\prod_i\delta_{k_i,k'_i}$. That is, we consider wavefunctions such that $\langle \vec{k},\bar{M}\vert \psi\rangle={ \sum_{\vec k}\int}\psi(M,\vec k) dM$, with $M$ belonging to the above mentioned interval. 

Taking into account equations \eqref{15} to \eqref{17}, and approximating\footnote{We did not find a closed form expression for the spectrum of $\left[\left(E^x_j\right)'\right]^2$, but the previous formula approximates it very well, up to corrections of the order $\Delta^2/x_0^2$, which are negligible for macroscopic black holes.} $\left[\left(E^x_j\right)'\right]^2$ by $\left(2 \sqrt{x_j^2+\Delta^2/x_0^2}+\delta x\right)^2$, the metric expectation values for the above mentioned physical state in Schwarzschild coordinates $F(x_j)=0$, are given by 
\begin{equation}
%ds^2=-\left(1-\frac{r_S}{\sqrt{x^2+x_0^2}}\right)dt^2+\frac{\left(x +\frac{\delta x}{2}\right)^2} {\left(x^2+x_0^2\right)\left(1-\frac{r_S}{\sqrt{x^2+x_0^2}}\right)}dx^2 +\left(x^2+x_0^2\right)\left(d\theta^2+\sin^2\theta d\varphi^2\right).
ds^2=-\left(1-\frac{r_S}{\sqrt{x^2+x_0^2}}\right)dt^2+\frac{\left(\sqrt{x^2+\Delta^2/x_0^2} +\frac{\delta x}{2}\right)^2} {\left(x^2+x_0^2\right)\left(1-\frac{r_S}{\sqrt{x^2+x_0^2}}\right)}dx^2 +\left(x^2+x_0^2\right)\left(d\theta^2+\sin^2\theta d\varphi^2\right).\label{18}
\end{equation}

In the previous expression we have dropped the label $j$ treating the variable $x$ as continuous in order to recover a smooth metric. However, it is possible to define a suitable discrete line element for the $(t,x)$ submanifold, as discussed in reference \cite{review}.

Computing the Einstein tensor of the metric (\ref{18}), we can get the effective stress energy tensor. Inside the trapped region, namely $0<x<\sqrt{r_S^2-x_0^2}$, at zeroth order in $\Delta/x_0$  it is determined by
\begin{eqnarray}\label{eq:rho}\nonumber
%\rho=-\frac{\left(\ell_{\rm Planck}+4 x_j\right)\ell_{\rm Planck}}{8\left(\ell_{\rm Planck}+2 x_j\right)^2 G \left(x_j^2+r_0^2\right)\pi}
\rho(x)&=&-\frac{G_{x}^x}{8\pi G}=\frac{1}{8\pi G}\frac{\delta x(4x+\delta  x)}{(\delta x+2x)^2(x^2+x_0^2)},\\ \nonumber
p_x(x) &=& \frac{G_{t}^t}{8\pi G}=-\frac{1}{8\pi G}\left(\frac{8\,\delta x \,r_S}{(\delta x+2x)^3\sqrt{x^2+x_0^2}}+\frac{\delta x(6\,x\,\delta x+\delta x^2 -8x_0^2 )}{(\delta x+2x)^3(x^2+x_0^2)}\right),\\
p_{||}(x) &=& \frac{G_{\theta}^\theta}{8\pi G}=\frac{1}{8\pi G}\left(2-\frac{r_S}{\sqrt{x^2+x_0^2}}\right)\frac{2\delta x}{(\delta x+2x)^3},
\end{eqnarray}
where $G_\mu^\nu$ are the components of the Einstein tensor. One can see that the components of this effective stress-energy tensor are smooth at the bounce and take their maximum value at the throat. If we set $\delta x = x_0$, and evaluate these quantities at the throat where the geometry ``bounces'', $x=0$, we obtain
\begin{eqnarray}\nonumber
\rho(x=0)&=&\frac{1}{8\pi G} \left(\frac{4\pi}{r_S\Delta}\right)^{2/3},\\\nonumber
p_x(x=0) &=&-\frac{1}{8\pi G}\left[\frac{32 \pi}{\Delta}-7\left(\frac{4\pi}{r_S\Delta}\right)^{2/3}\right],\\ 
p_{||}(x=0) &=&-\frac{1}{8\pi G}\left[\frac{8\pi}{\Delta}-4\left(\frac{4\pi}{r_S\Delta}\right)^{2/3}\right].
\end{eqnarray}
The pressures of the effective anisotropic fluid take universal (mass independent) values in the limit $r_S\gg \sqrt{\Delta}$. The space-time metric is also smooth at the bounce, something that can be explicitly checked by computing its spatial ``derivative''  (since one has a discrete space it is really a finite difference) at both sides of it. Moreover, note that both $p_x(x=0)$ and $p_{||}(x=0)$ are negative and order Planck while $\rho(x=0)$ is negligible (in the limit $r_S\gg \sqrt{\Delta}$), therefore we conclude that the null energy condition is violated \cite{nec}. 

Even though the above considered gauge fixing is the simplest, 
the effective metric is not well defined at the horizon. However, one can introduce a different gauge choice $F(x_j)$ associated with penetrating coordinates ---for instance the Painlevé--Gullstrand ones discussed before---, and work with a regular metric instead.

It should be noted that different elements of the physical space of states $\vert \vec k,M\rangle$, when quantum effects are taken into account, correspond to different physical geometries. However all these solutions have the same classical continuum limit when $\hbar \to 0$ and reproduce the Schwarzschild black hole.

\section{Conclusions}

Summarizing, we have shown that loop quantum gravity applied to spherically symmetric space-times contains quantum states that can lead
 naturally to the Simpson--Visser traversable wormhole with the throat inside a horizon. The radius of the throat is Planck scale, but with a large factor in front proportional to the ADM mass of the space-time  elevated to the $1/3$ power. The solution is free of shells of negative mass. This opens the possibility of studying the stability of the effective semiclassical solutions close to the singularity under gravitational or matter perturbations.  The semi-classical solutions have effective matter in them that violates energy conditions, that justifies that the singularity is resolved, but do not contain discontinuities that would lead to the emergence of shells of matter as had been encountered in previous solutions. The construction illustrates the freedom that one has in specifying the quantum states in vacuum spherically symmetric quantum gravity. Other interesting semiclassical solutions might also be possible. One may speculate that in the full theory the quantum states will be more constrained.
 
 \section{Acknowledgments}
This work was supported in part by Grant NSF-PHY-1903799, NSF-PHY-2206557, funds of the
Hearne Institute for Theoretical Physics, CCT-LSU, Pedeciba, Fondo Clemente Estable
FCE 1 2019 1 155865 and the Spanish Government through the projects  PID2020-118159GB-C43, PID2019-105943GB-I00 (with FEDER contribution), and the ``Operative Program FEDER2014-2020 Junta de Andaluc\'ia-Consejer\'ia de Econom\'ia y Conocimiento'' under project E-FQM-262-UGR18 by Universidad de Granada.


\begin{thebibliography}{99}
\bibitem{review} 
R.~Gambini, J.~Olmedo and J.~Pullin,
``Quantum geometry and black holes,''
[arXiv:2211.05621 [gr-qc]], to appear in "Handbook of Quantum Gravity", Cosimo Bambi, Leonardo Modesto, Ilya Shapiro (editors), Springer (2023) and references therein.

\bibitem{lqc} 
See for instance I. Agullo, P. Singh in “Loop quantum gravity: the first ´
30 years”, A. Ashtekar, J. Pullin (editors), World Scientific, Singapore
(2017) and references therein.

\bibitem{fireworks}
R. Cowen,  
%Quantum bounce could make black holes explode. 
Nature (2014) https://doi.org/10.1038/nature.2014.15573.

\bibitem{shells}
R. Gleiser (private communication); Y.~C.~Liu, J.~X.~Feng, F.~W.~Shu and A.~Wang,
%``Extended geometry of Gambini-Olmedo-Pullin polymer black hole and its quasinormal spectrum,''
Phys. Rev. D \textbf{104}, no.10, 106001 (2021)
doi:10.1103/PhysRevD.104.106001
[arXiv:2109.02861 [gr-qc]].

\bibitem{improved}
R.~Gambini, J.~Olmedo and J.~Pullin,
%``Loop Quantum Black Hole Extensions Within the Improved Dynamics,''
Front. Astron. Space Sci. \textbf{8}, 74 (2021)
doi:10.3389/fspas.2021.647241
[arXiv:2012.14212 [gr-qc]];
%``Spherically symmetric loop quantum gravity: analysis of improved dynamics,''
Class. Quant. Grav. \textbf{37}, no.20, 205012 (2020)
doi:10.1088/1361-6382/aba842
[arXiv:2006.01513 [gr-qc]].

\bibitem{SimpsonVisser}
A.~Simpson and M.~Visser,
%``Black-bounce to traversable wormhole,''
JCAP \textbf{02}, 042 (2019)
doi:10.1088/1475-7516/2019/02/042
[arXiv:1812.07114 [gr-qc]].



\bibitem{MorrisThorne}
M. S. Morris and K. S. Thorne, %“Wormholes in space-time and their use for interstellar
%travel: A tool for teaching general relativity”, 
Am. J. Phys. 56 (1988) 395.
doi:10.1119/1.15620

\bibitem{towards}
R.~Gambini, J.~Olmedo and J.~Pullin,
%``Towards a quantum notion of covariance in spherically symmetric loop quantum gravity,''
Phys. Rev. D \textbf{105}, no.2, 2 (2022)
doi:10.1103/PhysRevD.105.026017
[arXiv:2201.01616 [gr-qc]].

\bibitem{nec} Ugur Camci
and Khalid Saifullah, 
%``Conformal Symmetries of the Energy–Momentum Tensor of Spherically Symmetric Static Spacetimes,''
Symmetry 14(4), 647 (2022).

\end{thebibliography}
\end{document}